# Impact of rare earth magnetic moment on ordering of Ru in $Sr_2RuREO_6$ (RE = Gd and Eu)

V.P.S. Awana[1,2,*], Rahul Tripaathi[1], V.K. Sharma[1], H. Kishan[1], E. Takayama-Muromachi[2] and I. Felner[3]

[1]*National Physical Laboratory, Dr. K.S. Krishnan Marg, New Delhi-110012, India*

[2]*Superconducting Materials Center, NIMS, 1-1 Namiki, Tsukuba, Ibaraki, 305-0044, Japan*

[3] *Racah Institute of Physics, The Hebrew University, Jerusalem, 91904, Israel*

We report synthesis and magnetization of $Sr_2RuREO_6$ (RE/Ru-211$O_6$) with RE = Gd and Eu. Both Gd,Eu/Ru-211$O_6$ are formed in distorted perovskite-type single phase with monoclinic $P2_1/n$ space group. Magnetization ($M/T$) measurements exhibited antiferromagnetic ordering of Ru moments for both Gd and Eu compounds at $T_N$ of around 30 K due to super-exchange Ru-O-O-Ru interaction. Interestingly enough, detailed *M*/H plots of Gd/Ru-211$O_6$ exhibited, the development of a ferromagnetic (*FM*) component at 20 K. Further at below 10 K, the *FM* component disappears and rather an spin freezing field ($H^{SF}$) of around 1000 Oe is seen at 5 K. In case of Eu/Ru-211$O_6$, the basic antiferromagnetic ordering of Ru moments remains invariant down to 5 K, and the appearance of *FM* component is not seen. It seems that the high magnetic moment of $8\mu_B$ for $Gd^{3+}$ influences the *AFM* ordering of Ru in case of Gd/Ru-211$O_6$.

* Corresponding Author: awana@mail.nplindia.ernet.in

## I. INTRODUCTION

The double perovskite $Sr_2RuREO_6$ (RE = rare earth) are studied extensively over the years [1,2]. Basically they are derived from $SrRuO_3$ with replacement of alternate Ru atoms by RE in the unit cell [3]. Alternate replacement of $Sr^{2+}/RE^{3+}$ results in two interesting effects i.e., increase of Ru valence from +4 to +5 and the change of ferromagnetic (*FM*) order (150 K) to anti-ferromagnetic (*AFM*) below 30 K. The ionic size difference between six fold coordinated $Ru^{5+}$ (0.565 Å) and various rare earths for example $Eu^{3+}$ (0.947 Å) is quite large and hence the *AFM* temperature (30 K) and exact ordered perovskite structure is not guaranteed for all the cases. In fact the *AFM* temperature seems to scale with RE/unit cell size [4]. Further it was reported earlier that the nature of *AFM* order for Ru in $Sr_2RuREO_6$ is not the same for all the RE, for example in case of Er some phase separation is reported with in *AFM* structure [3]. Despite the fact that well documented neutron scattering experiments are available, the influence of RE on Ru AFM order in $Sr_2RuREO_6$ is not yet studied in detail [3]. In this direction a very recent study on $Sr_2RuGdO_6$ reported an FM component below AFM ordering due to possible Ru-O-Gd interactions [4].

Furthermore the possible influence of RE moment on Ru magnetic ordering in rutheno-cuprate magneto-superconductors $RuSr_2GdCu_2O_8$ and $RuSr_2(Gd,Ce)Cu_2O_{10}$ [5,6] has increased the interest in various ruthenates containing RE. Keeping in mind the very recent report on $Sr_2RuGdO_6$ [4] and our continuous interest on various rare earths containing rutheno-cuprates [7], we studied the magnetization of both magnetic RE containing $Sr_2RuGdO_6$ and non-magnetic RE based $Sr_2EuGdO_6$. Our choice of Eu and Gd guarantees the closest possible neighbors in terms of their ionic size [3]. These compounds are formed in an essentially single-phase form, without any trace of *FM* (150 K) phase of $SrRuO_3$. It is concluded on the basis of detailed magnetization measurements, that RE magnetic moment influences the nature of Ru magnetic ordering by Ru-O-RE interactions.

## II. EXPERIMENTAL DETAILS

Samples of composition $Sr_2RuREO_6$ with RE = Gd and Eu were synthesized through a solid-state reaction route. Calcinations were carried out on the mixed powder at 1000, 1100, 1200 and 1350 $^0$C each for 24 hours with intermediate grindings. The pressed bar-shaped pellets were annealed in a flow of oxygen at 1000 $^0$C for 40 hours and subsequently cooled slowly over a span of another 20 hours down to room temperature X-ray powder diffraction patterns were obtained by a diffractometer with Cu $K_\alpha$ radiation. DC susceptibility data were collected by a SQUID magnetometer (Quantum Design, MPMS).

## III. RESULTS AND DISCUSSION

The X-ray diffraction patterns of the $Sr_2RuREO_6$ with RE = Gd and Eu are depicted in Fig. 1. The powder diffraction pattern resembles that of other iso-structural compounds reported in the literature [1,3,4]. The entire pattern can be indexed on the basis of a distorted perovskite-type crystal structure in monoclinic $P2_1/n$ space group. Within the sensitivity limit (~3%) of the X-ray machine, the phase purity of the studied compounds is assured. In particular the parent *FM* $SrRuO_3$ phase is not available in these samples. The lattice parameters are $a$ = 5.793(2) Å, $b$ = 5.832(1) Å, $c$ = 8.2134 (7) Å and β = 90.31(1) for Gd and $a$ = 5.811(4) Å, $b$ = 5.846(3) Å, $c$ = 8.2269(4) Å and β = 90.35(1) for Eu, which is in good agreement with earlier reports [3,4].

Fig. 2 depicts the magnetization (*M*) versus temperature (*T*) plot for $Sr_2RuGdO_6$ sample in field-cooled (*FC*) situation under an applied field of 1000 Oe. As is evident from this figure, the sample is paramagnetic down to around 32 K with a small saturation/down turn in *M*/*T* near this temperature, with a further shoot up and a clear *FM* like saturation between 25 K and 10 K. Below 10 K, there is a further shoot up in the magnetization down to 5 K. Though *M*/*T* measurements generally do not lend to exact magnetic structure of a compound, they certainly provide an indication for the same. It seems from the plot in Fig. 2, that $Sr_2RuGdO_6$ undergoes an *AFM* like transition at 32 K and develops a *FM* component below 25 K to 10 K, with a further super-paramagnetic like behavior down to 5 K. As the ordering temperatures discussed above seem too high for the ordering of Gd moments, the ordering above 5 K is mainly due to Ru moments. In general the *M/T* behavior of presently studied $Sr_2RuGdO_6$ sample is in accordance with a very recent report on this compound [4]. The only difference is that our sample exhibited clear *FM* like saturation of the moments in *M*/*T*) between 25 K and 10 K. We also carried out low field *FC* and *ZFC* (zero-field-cooled) *M*/*T* measurements (plots not shown) to check the presence of $SrRuO_3$, which is ferromagnetic at above 150 K. In our sample the *FC* and *ZFC* branching is not observed at higher *T* above 32 K, hence we conclude that our sample is free of $SrRuO_3$ impurity.

For comparison with non magnetic RE ion containing $Sr_2RuREO_6$ system, in inset of Fig. 2 we show the *M*/*T* behavior of RE = Eu compound in field-cooled (*FC*) situation under an applied field of 1000 Oe. Interestingly $Sr_2RuEuO_6$ sample exhibits only the *AFM* ordering at around 28 K and no further FM or paramagnetic like situation is observed down to 5 K. It seems in case of non-magnetic RE ion Eu, the AFM ordering of Ru spins at around 28 K is not affected. This is in general agreement with earlier reports on $Sr_2RuREO_6$ system [1-3,8].

Low field *M*/*T* behaviors of $Sr_2RuGdO_6$ and $Sr_2RuREO_6$ systems in applied fields of 10 and 40 Oe respectively are shown in Fig. 3. It is generally believed that various minor magnetic transitions/anomalies gets wiped out in higher applied fields. The *ZFC M*/*T* plot (Fig 3) of $Sr_2RuGdO_6$ in applied field of 10 Oe is very similar to that as observed for the same sample under magnetic field of 1000 Oe (Fig 2). In particular all the magnetic transitions are seen in terms of an *AFM* like transition at 32 K, *FM* component below 25 K to 10 K, and super-paramagnetic like behavior down to 4K. Further an *AFM* transition presumably due to $T_N$ (Gd) is seen below around 3 K. Above 32 K the system is purely paramagnetic.

The M/*T* behavior of $Sr_2RuREO_6$ system in an applied field of 40 Oe and in *ZFC* situation is depicted in inset of Fig.3. Clear *AFM* like transition, originating from Ru spins is seen below 28 K. This compound remains *AFM* down to 5 K, and no other magnetic anomalies are visible. Interestingly the magnetic susceptibility above 28 K does not appear to be purely paramagnetic. Detailed fitting of the magnetic susceptibility for $Sr_2RuREO_6$ system is reported in ref. [3]. In general Eu is in trivalent non-magnetic state and possess large amount of temperature independent contribution to the paramagnetic susceptibility of Ru in $Sr_2RuREO_6$ system [3]. In our recent paper (ref. 8) we measured the Mossbauer spectra of $Sr_2RuREO_6$ system at 90 and 4.2 K. The well-defined single magnetic spectrum at 4.2 K indicated the exchange field induced on the Eu ions by the Ru sheets. The estimated induces magnetic moment on the Eu ions is: $0.35\mu_B$. The fact that the negative effective quadrupole interaction at 4.2 K is just half of the quadrupole interaction at 90 K indicates clearly that the induced magnetic hyperfine field is along the c –axis (Type I AFM) as stated by Cao et. al [9] on $Sr_2RuYO_6$. Combining the results in ref. 3, 4, 8, 9 and the present outcome, one can say that these compounds possess complex magnetic structure in case of magnetic rare earths and Y. In particular the *AFM* ordering of Ru spins turn into canted *FM* and super paramagnetic behavior in case of Y [9] and magnetic rare earths [3,4,8]. There is an urgency to revisit the neutron scattering experimentations on the $Sr_2RuY/REO_6$ compounds. Though the magnetic structure is mainly due to the *AFM* ordering of Ru spins below say 30 K, the detailed magnetization and Mossbauer spectroscopy under various applied fields do indicate towards canting of *AFM* moments and appearance of *FM/Spin-glass* structure [9].

To further elucidate the magnetism of $Sr_2RuREO_6$ system, we show the *M*(*H*) plots of both RE = Gd and Eu compounds at close temperature intervals up to applied fields of 5 Tesla in Figs 4-6. Fig. 4 shows the *M/H* plot for $Sr_2RuGdO_6$ at 5 K with $H = \pm 5$ Tesla. Though no loop opening is seen, a spin flop like situation occurs at around $H = \pm 1$ Tesla, which is marked as $T^{SF}$. A similar behavior is shown very recently in ref. 4. The *M/H* plot for $Sr_2RuEuO_6$ at 5 K with $H = \pm 5$ Tesla is shown in inset of Fig.4, which is linear with field. This shows that $Sr_2RuEuO_6$ is *AFM* at 5 K and no indication of any spin flop like situation is seen. Clearly the magnetic structure of Ru spins in non-magnetic RE containing $Sr_2RuEuO_6$ is different than that of magnetic RE based $Sr_2RuGdO_6$. The *M/H* plots of both RE = Gd and Eu based $Sr_2RuREO_6$ system at 15 K are shown in Fig. 5. The situation at 15 K is very similar to that as at 5 K, in terms of $T^{SF}$ for Gd (main panel, Fig. 5), and *AFM* like structure for Eu compound (inset II, Fig.4), the important difference is the opening of *M/H* loop for $Sr_2RuGdO_6$ (upper inset I, Fig.4). It seems that $Sr_2RuEuO_6$ is *AFM*, the $Sr_2RuGdO_6$ has a *FM* component. To further elucidate the *FM* component of $Sr_2RuGdO_6$ system, we show in Fig. 6 the low field *M/H* plots of this system. Clear *M/H* loop opening at 20 K for $Sr_2RuGdO_6$ system indicates towards the *FM* component in this system. On the other hand the *M/H* plot for $Sr_2RuEuO_6$ at 20 K is completely linear (inset Fig.6). Interestingly at 30 and 40 K the *M/H* data for $Sr_2RuGdO_6$ and $Sr_2RuEuO_6$ systems is linear, without any *FM* like hysteresis/opening, plots not shown. Summarily one can say that, though non-magnetic rare earth containing $Sr_2RuEuO_6$ system remains *AFM* below 30 K, in $Sr_2RuGdO_6$ system the *AFM* ordered Ru spins at 30 K develops a *FM* component with in the *AFM* arrangement. The *AFM* ordering of Ru spins in $Sr_2RuREO_6$ systems below 30 K is known to be due to Ru-O-O-Ru coupling [1-3,4]. In case of $Sr_2RuGdO_6$ system the *AFM* ordering of Ru spins seems to develop a *FM* component through possible Ru-O-RE interactions below 20 K due to magnetic RE (Gd).

In conclusion, we can safely conclude that the high magnetic moment of $8\mu_B$ for $Gd^{3+}$ influences the *AFM* ordering of Ru in case of Gd/Ru-$211O_6$, which is obviously absent in case of non- magnetic RE containing Eu/Ru-$211O_6$. There is a possibility that Ru-O-RE interactions take place along with Ru-O-O-Ru coupling and hence in case of magnetic RE (Gd), the *AFM* ordering of Ru is influenced, but not in case of non-magnetic Eu.

This work is partially supported by INSA-JSPS bilateral exchange visit of Dr. V. P. S. Awana to NIMS Japan. This research is also supported by the INDO-ISRAEL collaborative DST (Department of Science and Technology, India) and MST (Ministry of Science and Technology, Israel) project. Authors from the NPL appreciate the interest and advice of Professor Vikram Kumar (Director) in the present work.

## FIGURE CAPTIONS

Fig. 1 X-Ray diffraction patterns of $Sr_2RuREO_6$ with RE = Gd and Eu

Fig. 2 Magnetization (*M*) versus temperature (*T*) plot for $Sr_2RuGdO_6$ sample in field-cooled (*FC*) situation under an applied field of 1000 Oe, inset exhibits the same for $Sr_2RuREO_6$ system

Fig. 3 Magnetization (*M*) versus temperature (*T*) plot for $Sr_2RuGdO_6$ sample in zero-field-cooled (*ZFC*) situation under an applied field of 10 Oe, inset exhibits the same for $Sr_2RuREO_6$ system at appled field of 40 Oe.

Fig. 4 *M*/*H* plot for $Sr_2RuGdO_6$ in applied fields of ± 5 Tesla at 5 K, inset exhibits the same for $Sr_2RuREO_6$ system

Fig. 5 *M*/*H* plot for $Sr_2RuGdO_6$ in applied fields of ± 5 Tesla at 15 K, inset I exhibits the same for low field of ± 3000 Oe. Inset II is the high field (± 5 Tesla) *M*/*H* plot for $Sr_2RuREO_6$ system at 15 K

Fig. 6 *M*/*H* plot for $Sr_2RuGdO_6$ in applied fields of ± 5000 Oe at 20 K, inset shows *M*/*H* plot for $Sr_2RuREO_6$ system at 20 K in applied fields of ± 5 Tesla

FIG.1.

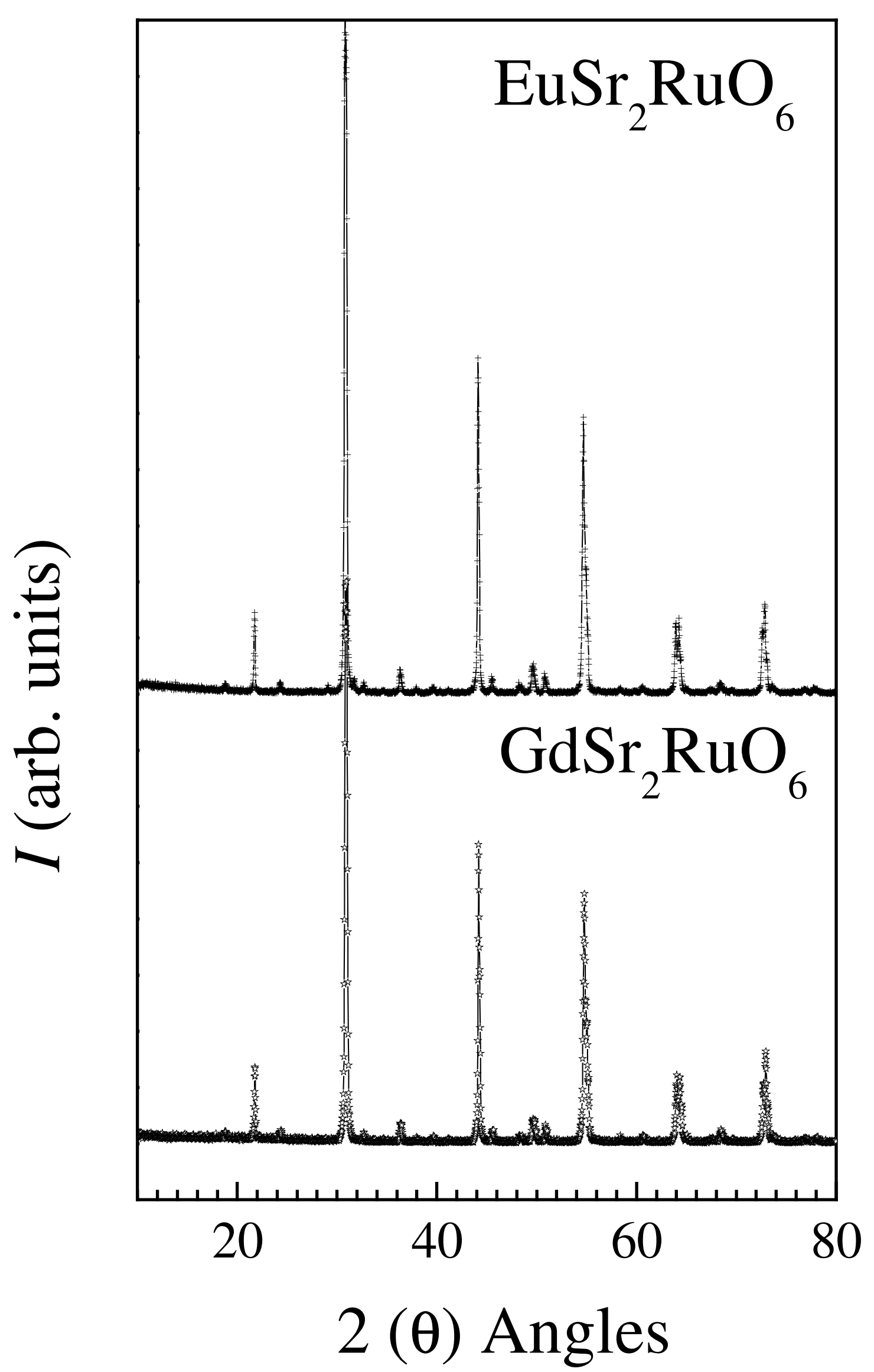

FIG.2.

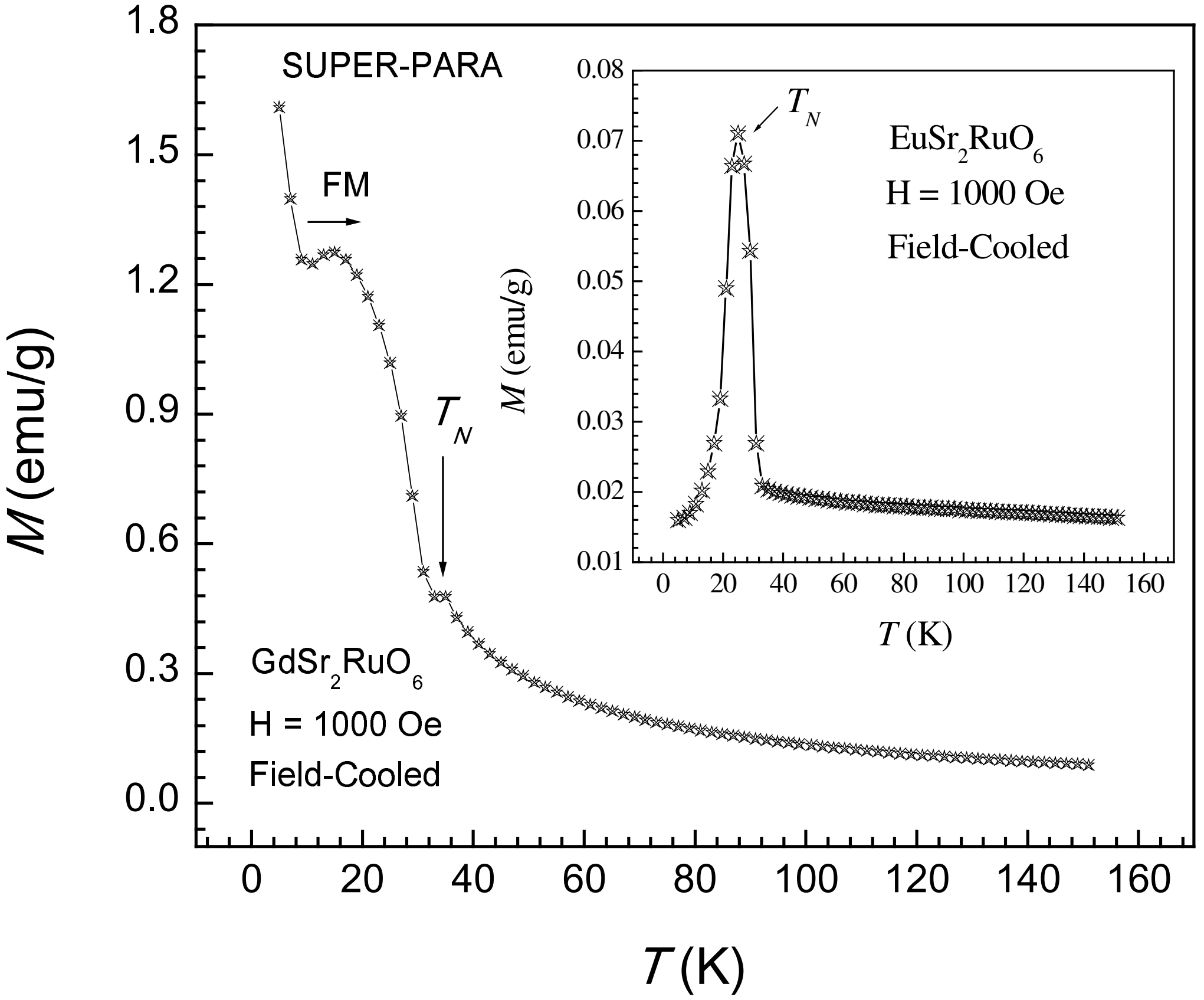

FIG.3.

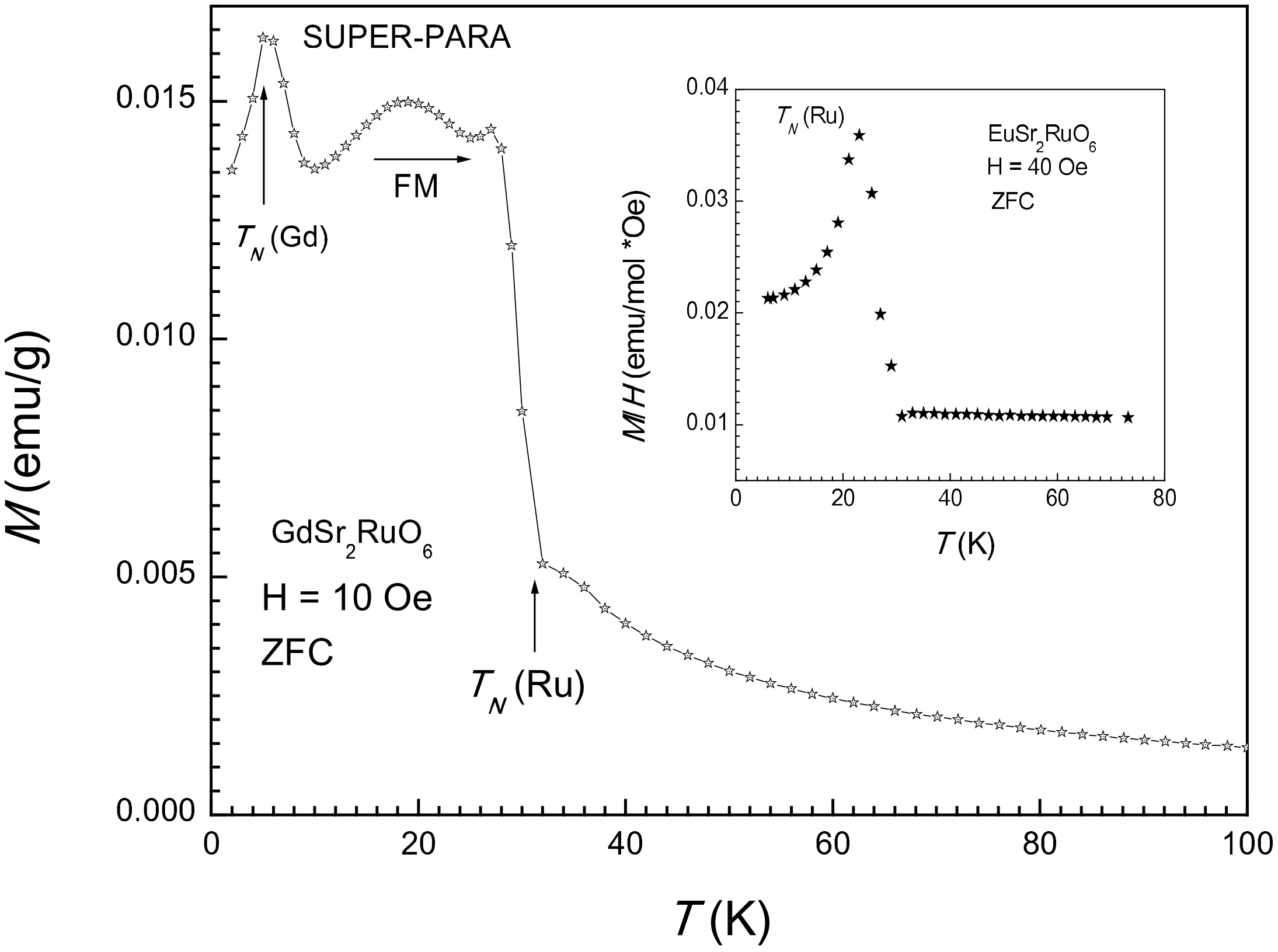

FIG.4.

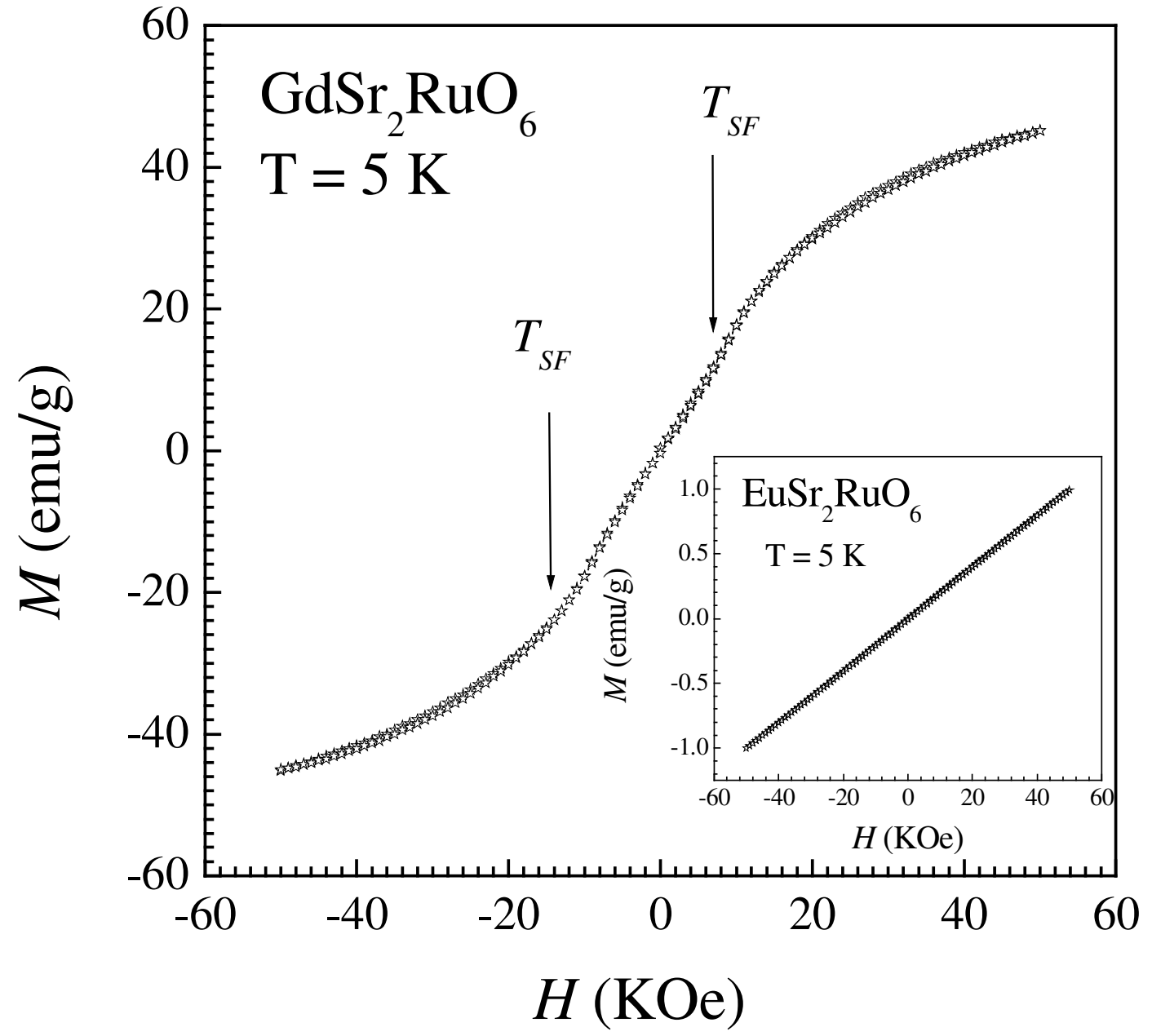

FIG.5.

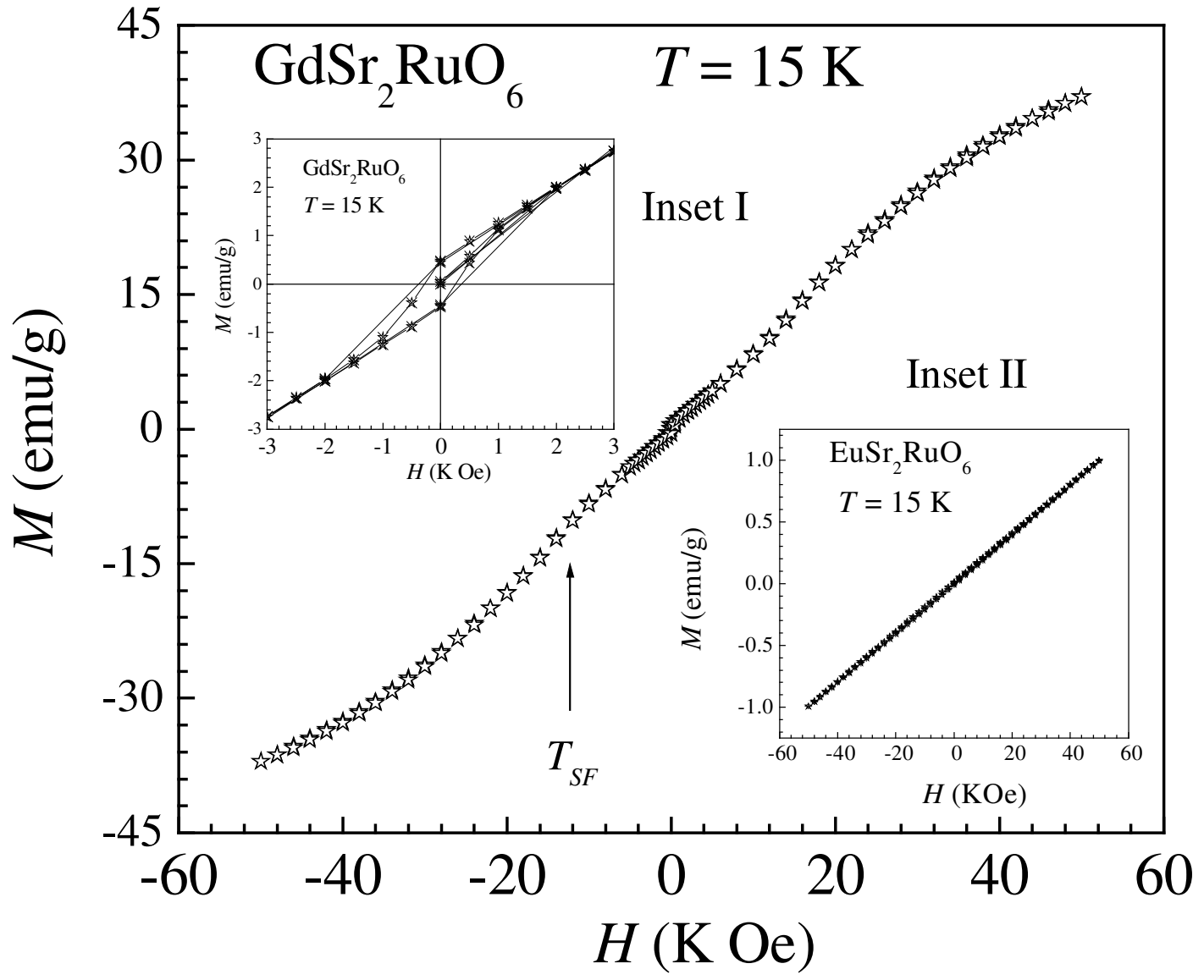

FIG.6.

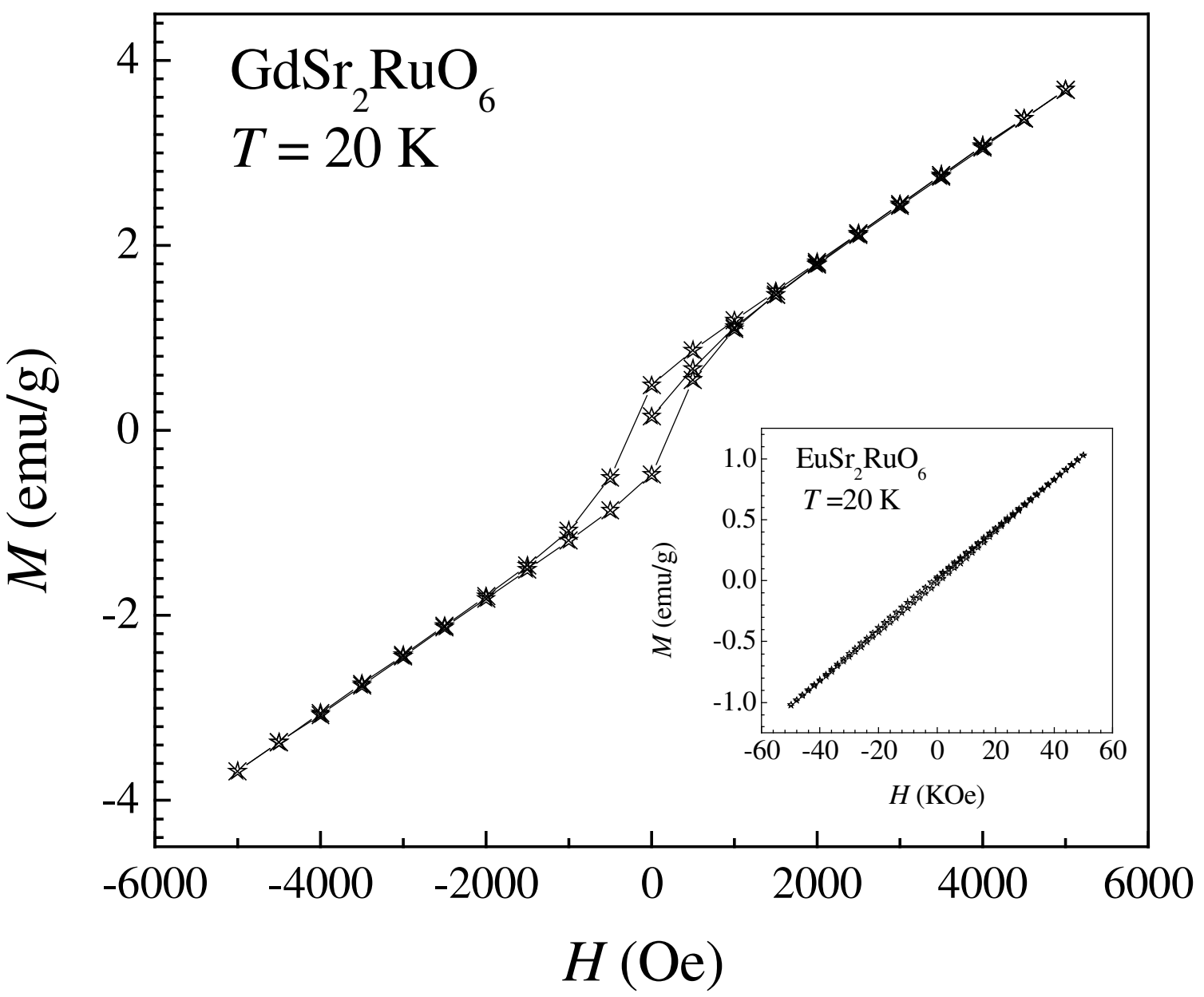